\documentclass[prl,twocolumn,amsmath]{revtex4-1}

\usepackage{amsmath}
\usepackage{amssymb}
\usepackage{color}
\usepackage[normalem]{ulem}
\usepackage{mathtools}
\usepackage{color}
\usepackage[justification=centerlast,font={small,singlespacing}]{caption}
\usepackage[justification=centerlast,font={small,singlespacing}]{subcaption}

\begin{document}
\title{Effective Potential for Cellular Size Control}
\author{David A. Kessler}
\email{kessler@dave.ph.biu.ac.il}
\author{Stanislav Burov}
\email{stasbur@gmail.com}
\affiliation{Physics Department, Bar-Ilan University, Ramat Gan 52900,
Israel}

\pacs{PACS}

\begin{abstract}
For various species of biological cells, experimental observations indicate the existence of universal distributions of the cellular size, scaling relations between the cell-size moments and simple rules for the cell-size control. 
We address a class of models for the control of cell division, and present the steady state distributions. By introducing concepts such as effective force and potential, we are able to address the appearance of scaling collapse of different distributions and the connection between various moments of the cell-size. Our approach allows us to derive strict bounds which a potential cell-size control scenario must meet in order to yield a steady state distribution.
The so-called ``adder" model for cell-size control exhibits the weakest control that still enables the existence of stable size distribution, a fact that might explain the relative ``popularity" of this scenario for different cells.
%We examine a class of models for the control of cell division, including the effects of stochasticity in the division time. We present the steady-state size distribution for this class of models.  We investigate in detail the timer, sizer and adder models. We show whereas the timer does not exhibit a steady-state distribution, the sizer is well-described by a log-normal distribution and the adder by a distribution with a power-law tail at large sizes.  We derive a stability condition for a steady-state distribution to exist.  We also demonstrate how scaling collapses generically occur is such models.
\end{abstract}

\maketitle

Probably the most  basic result of equilibrium statistical physics is that knowledge of the potential energy function and the temperature are sufficient to completely describe a system's long-time behavior~\cite{FeynmanStatistic}. The function $\exp\left(-U(x)/k_B T\right)$ is the only ingredient of the Boltzmann distribution, and a knowledge of the energy ($U$) in any given state ($x$) of the system (for a given temperature $T$) determines the distribution of physical observables in equilibrium. The challenge of finding similar approaches for out of equilibrium systems has fascinated the scientific community and many general advances in this field have been made. One of the most important examples of an out of equilibrium system is a living biological system.
There are many differences between live biological systems and the physical systems that are usually addressed in statistical mechanics. The tremendous  complexity in biology makes the idea of existence of a single concept (like energy) that determines the observed rich behavior look like a non-realistic pipe dream.   
Nonetheless, the  presence of significant levels of noise together with the known cases where simple statistical behavior  emerges out of the many underlying processes, gives hope for the  possible application of statistical physics to specific properties of living matter~\cite{Popkin}. 
One such property is  cellular growth and division, which has been addressed in many recent experimental studies~\cite{Giometto, Siggia,Tanuchi,Balaban,Tans}. For protein expression in bacteria a connection between the first and second moments of cellular size was detected~\cite{Xie, HSalman, Naama1} (a relation also known as Taylor's law). The size of several unicellular eukaryotes  was observed to attain a stable distribution, rescalable by the first two moments~\cite{Giometto}. Similar behavior was observed for the protein number~\cite{Naama1} and cell size~\cite{Kennard} in {\it E. coli}.  The distribution of inter-division times of {\it C. crescentus} were also seen to exhibit a scaling collapse~\cite{Sri}. Simple rules that govern cellular growth were verified experimentally for several bacteria types~\cite{Campos,SookJoon,Soifer}. For example, in {\it E. coli} it was shown that the devision occurs when the bacteria grows by a constant amount~\cite{Campos,SookJoon,Soifer}, a scenario termed as ``adder" dynamics~\cite{Amir1,Taheri}. This model was explored theoretically~\cite{Amir1} and (under specific simplifications) shown to fit beautifully the observed stable size distribution of the bacteria~\cite{Soifer}. The protein number during the bacterial growth was successfully described by a model sharing similar principles~\cite{Naama1}. 

Our goal in this manuscript is to address the connection between a given growth model and the emergent steady state distribution. We will show that all the abovementioned behaviors of rescalable distributions, connection between moments, i.e., Taylor's law, and stability criteria, can be described in terms similar to energy and temperature in statistical physics. We will derive an effective potential for cellular growth and show how the notion of effective temperature is different for various growth models. The relatively frequent appearance of the ``adder" model will become clear when addressed from the standpoint of the weakest form of the potential that produces a stable distribution.  Moreover, we show how the same criteria that gives rise to the adder model predicts the appearance of broad power-law tails for the stable distribution. 
  
 For the growth of a bacteria, there is a growing number of experimental studies that clearly show an exponential growth of the bacteria volume till its division. At division, the bacteria splits into two parts, of equal size for symmetric division and otherwise for asymmetric division.  We will restrict ourselves in this study to symmetric division~\cite{Amir2}. We define $v_n$ to be the size of the bacteria after the $n$'th division. The size at which the bacteria will divide  the next time is then $2v_{n+1}$.  In general, a growth law states that this size $2v_{n+1}$ is dependent on the size of the cell at the beginning of the cycle, i.e. $2v_{n+1}=f(v_n)$, where $f(\cdot)$ is some specified function. The time difference between the $n$'th and $(n+1)$'th division is given by $t_{n+1}=\tau\left[ \ln\left( f(v_{n})/v_{n} \right) +\eta_{n}\right]$. For simplicity, the fluctuations in the growth rate $1/\tau$ are neglected and the term $\eta_{n}$ is the temporal noise accumulated through the growth process. There are several competing scenarios for the growth laws, i.e. $f(\cdot)$, that exist in the literature~\cite{BacteriaBook1,BacteriaBook2,adder01,SookJoon,Amir1}. The models that have attracted the most interest are: {\it(i)} The "timer": the cell grows for a specific  amount of time (up to noisy fluctuations). In this model $f(v_n)=\theta v_n$, such that the average growth time is $\tau\ln(\theta)$. {\it(ii)} The``sizer": the cell grows till a specific size $C$. Here $f(v_n)=C$. {\it(iii)} The ``adder": a specific amount of volume/mass is added through the growth process. $f(v_n)=v_n+\Delta$. The specific growth law for a specific species of bacteria must be inferred from experiment, although it is not always an obvious task~\cite{Taheri,Mixer}.

 Our basic model for the variation of cell size with generation number $n$ is then the stochastic map (SM)
 %%%%%%%%%%%%%%%%%%%%%%%
 \begin{equation}
 v_{n+1}=\frac{1}{2}v_{n}\exp\left[ \ln\left( f(v_{n})/v_{n} \right) +\eta_{n}\right].
 \label{basic01}
 \end{equation}
 %%%%%%%%%%%%%%%%%%%%%%%
It is more convenient to define $a_n\equiv \ln(v_n)$ and then Eq.~(\ref{basic01}) takes the simple form
%%%%%%%%%%%%%%%%%%%%%%%
 \begin{equation}
 a_{n+1}= a_n+g(a_n)+\eta_{n},
 \label{basic02}
 \end{equation}
 %%%%%%%%%%%%%%%%%%%%%%%`\
where $g(a_n)=\ln\left[f(\exp(a_n))/2\right]-a_n$. Due to the explicit relation between $a_n$ and the cell size $v_n$, it sufficient to determine the statistical behavior of $a_n$.
 In order to obtain a criteria for determining whether a model exhibits  control of bacterial size, it is sufficient to show that $a_n$ attains a stable distribution as the generation number $n$ grows. It is then a question of the stability criteria for the map Eq. \ref{basic02},  where the function $g(\cdot)$ can be quite nonlinear. 
The criteria for stability immediately rules out any growth laws that produce a $g(\cdot)$ that diverges too fast as $a_n\to\pm\infty$. Indeed, if $g(a_n)\sim a_{n}^2$,  (when $|a_n|$ is sufficiently large) $a_n$ grows in an unbounded fashion with $n$ and no stable distribution will be obtained.  Eq.~(\ref{basic02}) can be viewed as a discretized form of  a continuous (in $n$) Langevin equation~\cite{Gardiner}, $da_n/dn=g(a_n)+\eta_n$. While such an approximation is not sufficient, it provides a general idea of the mathematical route we intend to follow. Indeed, a Langevin equation with $\delta$ correlated noise describes a particle in a bath with specific temperature under the affect of an external force~\cite{Gardiner,Risken}. 
The advantage of such a description of the cellular division and growth process lies in the efficient mathematical tools that were derived in order to describe the equilibrium behavior of a system. The bacterial case is a nonequilibrium one, but the size control procedure eventually creates a population with a stable size distribution. This stable distribution will be achieved by means similar to how the equilibrium distribution is obtained for a thermal (nonliving) system.

 %%%%%%%%%%%%%%%%%%%%%%%%%%%%%%%%%%%%%%%%%%%%%%%%%
\begin{figure}[h!]
\begin{center}

                \includegraphics[width=0.45\textwidth]{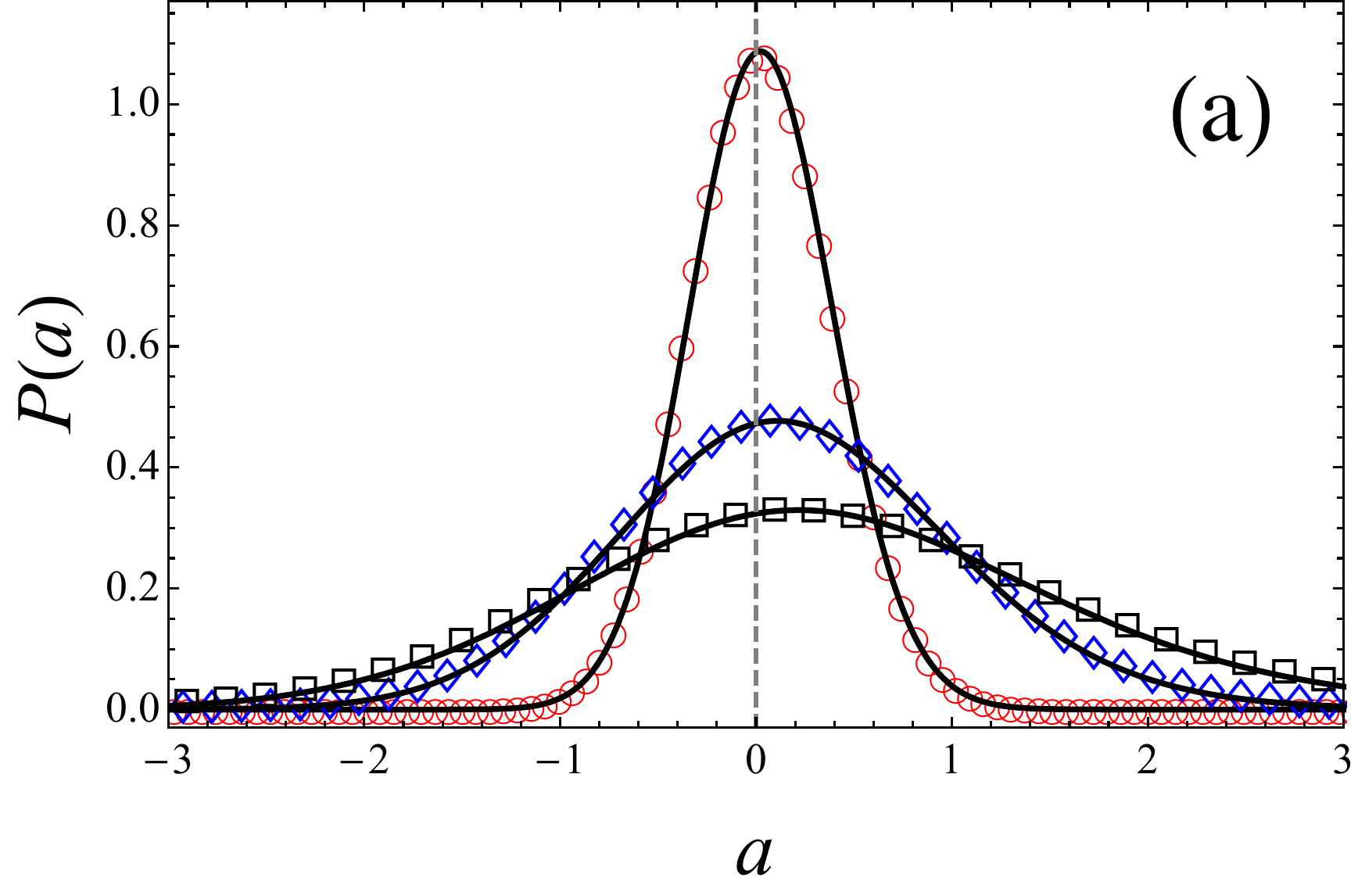}

                \includegraphics[width=0.45\textwidth]{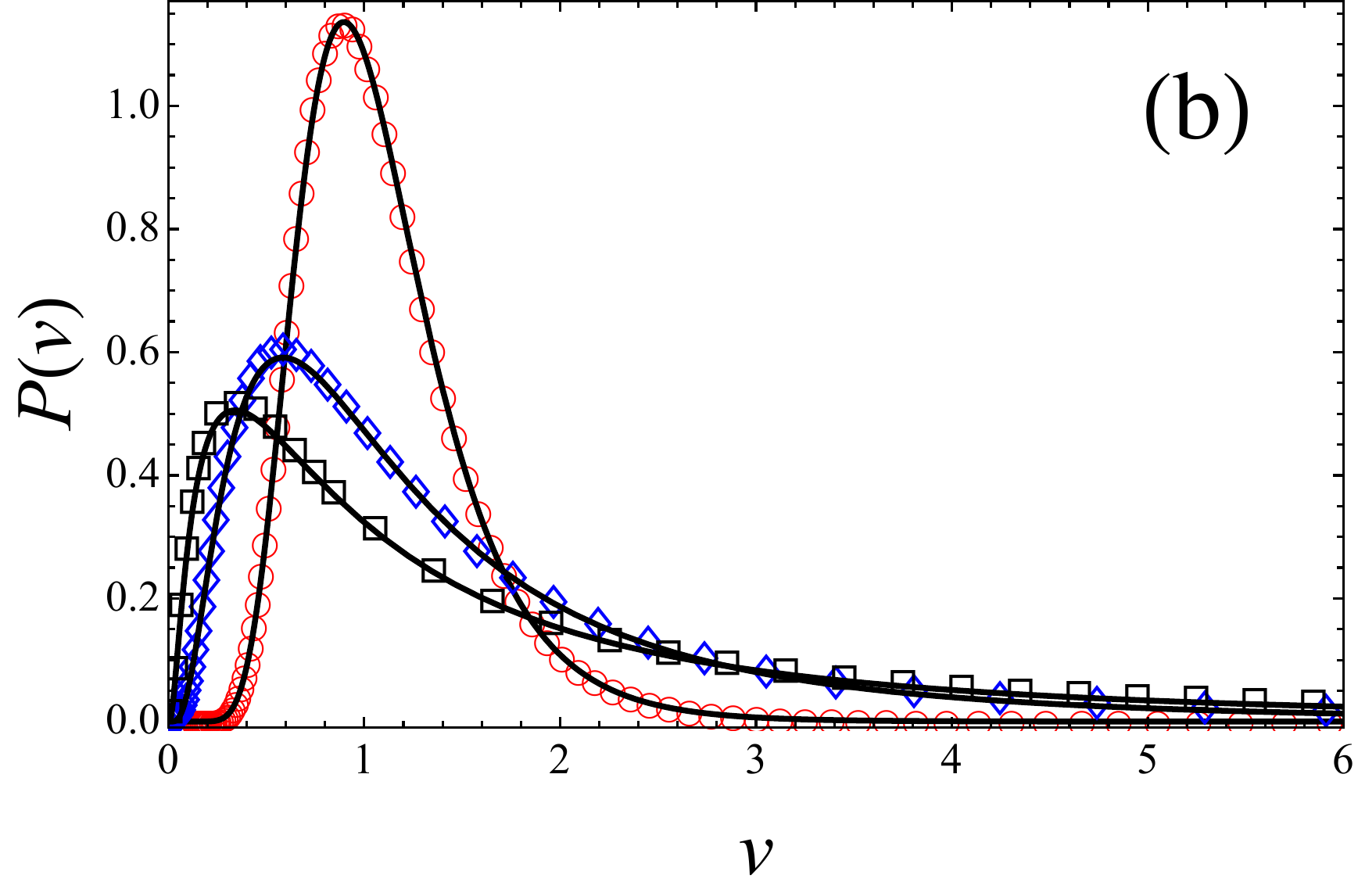}

                \includegraphics[width=0.45\textwidth]{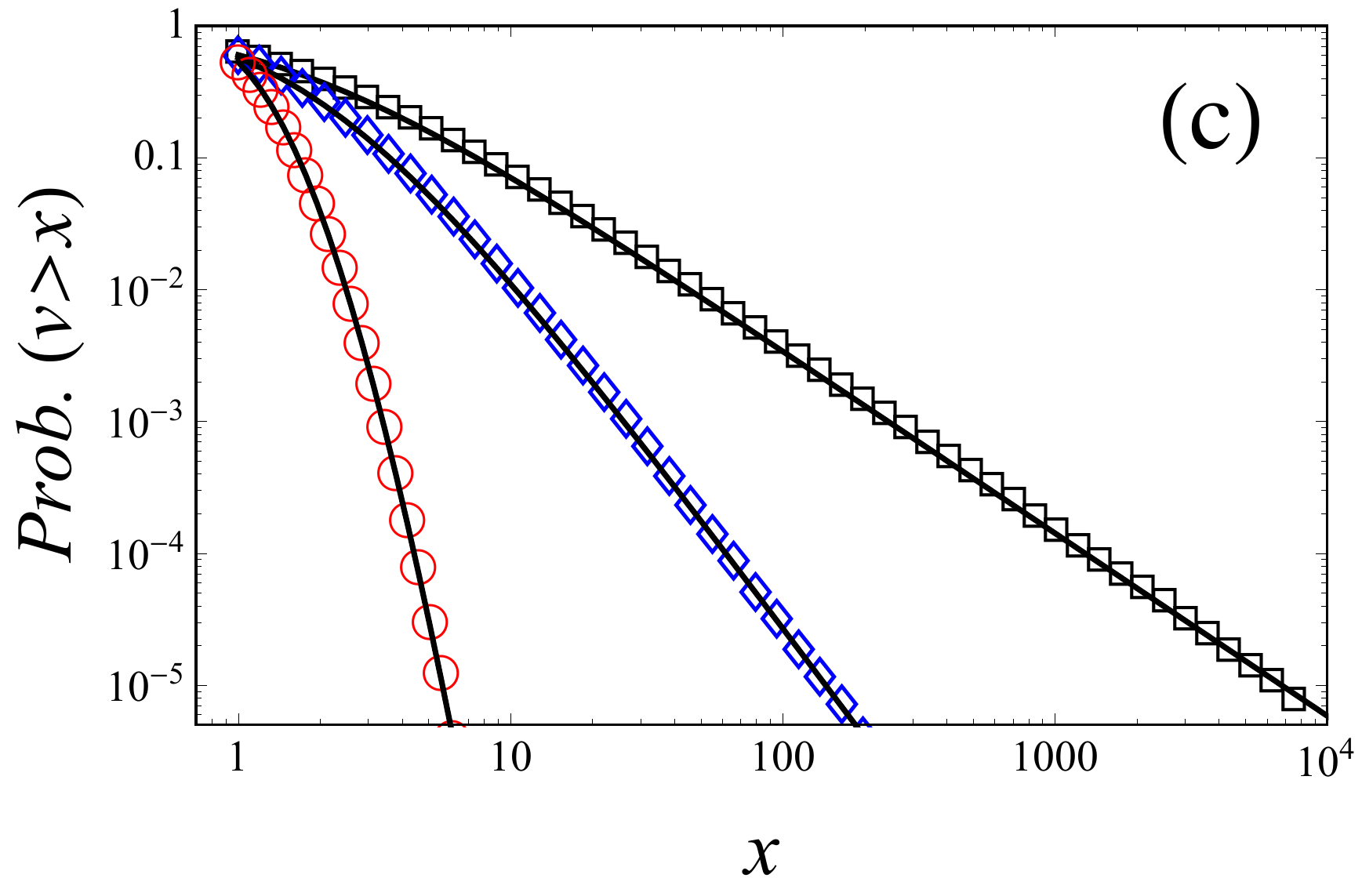}    
\end{center}
\caption{ Distributions of cell sizes $v$ and the appropriate variable in log-space $a=\ln\left(v\right) $ for the adder model. Three different noise strengths $\langle \eta^2 \rangle$ were used: $\langle \eta^2 \rangle=0.1$ (circles \textcolor{red}{$\bigcirc$} ),  $\langle \eta^2 \rangle=0.1$ (dimonds \textcolor{blue}{$\Diamond$} ) and $\langle \eta^2 \rangle=0.1$ (squares \textcolor{black}{ $\Box$} ). The parameter $\Delta$ is $1$ in all plots. Panel {\bf(a)} displays the distributions of $a$, panel {\bf(b)} displays the distributions of $v$. Thick lines are appropriate analytic approximation by the virtue of Eqs.~(\ref{boltzmn}) and (\ref{adderEnrg}). Panel {\bf(c)} displays the asymptotic power-law decay of the $v$ distributions. The decay is consistent  with Eq.~(\ref{powerlaw}). 
}
\label{figcomparing}
\end{figure}
%%%%%%%%%%%%%%%%%%%%%%%%%%%%%%%%%%%%%%%%%%%%%%%%%

In recent study~\cite{KesBur} we derived a general approach for deriving a continuous approximation for SM's of the form Eq. (\ref{basic02}). The so-called second order approximation for Eq.~(\ref{basic02}) is a Langevin Equation with multiplicative noise 
%%%%%%%%%%%%%%%%%%%%%%%%
\begin{equation}
\begin{split}
da_n = \frac{g\left(a_n\right)-
\frac{1}{4}\frac{\partial^2 g\left(a_n\right)}{\partial a_n^2}
\left[\frac{\sqrt{\langle\eta^2\rangle}}{1+\frac{1}{2}\frac{\partial g(a_n)}{\partial a_n}}  \right]^2}
{1+\frac{1}{2}\frac{\partial g(a_n)}{\partial a_n}}dn
+
\frac{\sqrt{\langle\eta^2\rangle}}{1+\frac{1}{2}\frac{\partial g(a_n)}{\partial a_n}}dB_n. 
\end{split}
\label{langevin}
\end{equation}
%%%%%%%%%%%%%%%%%%%%%%%%%%%
The generation number $n$ is now treated as a continuous parameter, $B_n$ is the Weiner process~\cite{Gardiner} (i.e., the noise term) and the brackets $\langle \dots \rangle$ represent the ensemble average value. Using the standard technique of deriving a Fokker-Planck equation for the distribution of $a_n$ \cite{Risken,Lau,KesBur},
the stable distribution $P(a)$ is of the form of a Boltzmann distribution
%%%%%%%%%%%%%%%%%%%%%%%%%%%%%%%%%%%%%%%%%
\begin{equation}
P(a)={\cal N}^{-1} \exp \left(-\frac{H(a)}{\langle \eta^2 \rangle}\right)
\label{boltzmn}
\end{equation}
%%%%%%%%%%%%%%%%%%%%%%%%%%%%%%%%%%%%%
where 
%%%%%%%%%%%%%%%%%%%%%%%%%%%%%%%%%%%%%
\begin{equation}
H(a) = -2\!\int \!g(a) da -\frac{1}{2}g^2(a)-\langle \eta^2 \rangle\ln\left(\left|1+\frac{1}{2}\frac{dg(a)}{da}\right| \right)
\label{energy}
\end{equation}
%%%%%%%%%%%%%%%%%%%%%%%%%%%%%%%%
plays the role of effective physical potential energy. ${\cal N}=\int_\infty^\infty\exp\left(-H(a)/\langle \eta^2 \rangle\right)\,da$ is the normalization constant. 
The analogy to physical situation is not perfect;  the noise strength $\langle \eta^2 \rangle$ plays the role of the temperature $k_b T$  but it also appears in $H(a)$, reflecting the presence of multiplicative noise~\cite{Lau,GitBur} in Eq.~(\ref{langevin}). 
Such a description of the effective potential energy allows the criteria for stability to be easily obtained: the potential energy must be binding. 
This must be used with caution since too strong a divergence of $H(a)$ will produce a situation where no stable distribution for $a$ exists, as previously mentioned. 
A simple example is the ``timer" model, where $g(a)=\ln(\theta/2)$ is a  constant. The function $exp\left(-H(a)/\langle \eta^2 \rangle\right)$ is non-normalizable and a stable distribution does not exist. Eq.~(\ref{langevin}) for such a case displays a behavior of a particle pushed in a specific direction by a constant force, a situation that can't attain positional equilibrium. The non-validity of the ``timer" as a possible size control scenario was already noted in Ref.~\cite{Amir1,Taheri}. 

We continue the exploration of possible scenarios. For the ``sizer", we have $g(a)=\ln(C/2)-a$ and so $H(a)=(\ln(C/2)-a)^2/2$, according to Eq.~(\ref{energy}). This is a case of a quadratic effective energy around the point $a=\ln(C/2)$. 
Our approximation for such case produces the Gaussian distribution $P(a)\sim\exp(-(\ln(C/2)-a)^2/2\langle \eta^2 \rangle)$, which is exact for the linear map~\cite{KesBur,McKane2,Amir1}. The appropriate distribution of the size $v$ is then log-normal, i.e., $P(v)\sim(1/v)\exp(-\left[\ln(2a/C)\right]^2/2\langle \eta^2 \rangle)$. 

The third scenario is the ``adder" where $g(a)=\ln\left[(\exp(a)+\Delta)/2\right]-a$ is here non-linear. The effective energy $H(a)$ for this case is
%%%%%%%%%%%%%%%%%%%%%%%%%%%%%%%%%%%%%%%
\begin{equation}
\begin{split}
H(a)&=\ln^2\left(2e^a/\Delta\right)+2 \text{Li}_2\left(-e^a/\Delta \right)
-\frac{1}{2}\ln^2 \left(\frac{e^a/\Delta+1}{2e^a /\Delta} \right)\\
&
-\langle \eta^2 \rangle \ln\left(\frac{2e^a/\Delta+1}{e^a/\Delta+1}\right)
%H(a)&=-\frac{1}{2} \left(\ln \left(\frac{e^a+\Delta}{2} \right)-a\right)^2
%+a \left(2 \ln \left(\frac{2 }{\Delta }\right)+a\right)
%\\
%&+2 \text{Li}_2\left(-\frac{e^a}{\Delta }\right)
%-\langle \eta^2 \rangle \ln\left(\frac{2\exp(a)+\Delta}{\exp(a)+\Delta}\right).
\end{split}
\label{adderEnrg}
\end{equation}
%%%%%%%%%%%%%%%%%%%%%%%%%%%%%%%%%%%%%%%%
Here, $\text{Li}_2\left(x\right)$ is the polylogarithm function~\cite{Erdely,polylog}. The asymptotic behavior of $H(a)$~\cite{Asymptotic} is $H(a)\sim a^2/2$ as $a\to-\infty$ and $H(a)\sim\ln(2)a$ as $a\to\infty$. This result can be simply explained by looking at the form of $g(a)$. Considering $g(a)$ as an effective force and using the effective relation between force and potential, as described by Eq.~(\ref{energy}), we obtain:  (i) for large negative $a$ the force is $\sim-a$, implying that the potential energy $\sim1/2 a^2$, (ii) for large and positive $a$ as $g(a)=-\ln(2)$ the potential energy should be  $\sim \ln(4) a$. This explains the observed Gaussian tail for negative $a$ and exponential decay for positive $a$. For the  size itself, $v$, the adder mechanism then predicts a power-law asymptotic behavior,
%%%%%%%%%%%%%%%%%%%%%%%%%%%%%%%%%%%%%%%%%%%%
\begin{equation}
P(v)\sim v^{-\left(1+\frac{\ln(4)}{\langle \eta^2 \rangle}\right)} \quad\text{for }\,\,v\to\infty.
\label{powerlaw}
\end{equation}
%%%%%%%%%%%%%%%%%%%%%%%%%%%%%%%%%%%%%%%%%%
The power law behavior, which is a direct consequence of exponential growth and the ``adder" scenario suggests the appearance of extremely large cells in the population. When $\langle \eta^2 \rangle\geq \ln(4)$ the average size of the cell diverges, meaning that in such a case we will encounter  bacteria that will grow for the whole time of the experiment (like the filamentous bacteria). Note that Ref. \cite{Amir1} approximated the ``adder" with a log-normal distribution, in contradiction to our Eq. (\ref{adderEnrg}), with its power-law tails. 

%%%%%%%%%%%%%%%%%%%%%%%%%%%%%%%%%%%%%%%%%%%%%%%%
\begin{figure}[h!]
\begin{center}
                \includegraphics[width=0.45\textwidth]{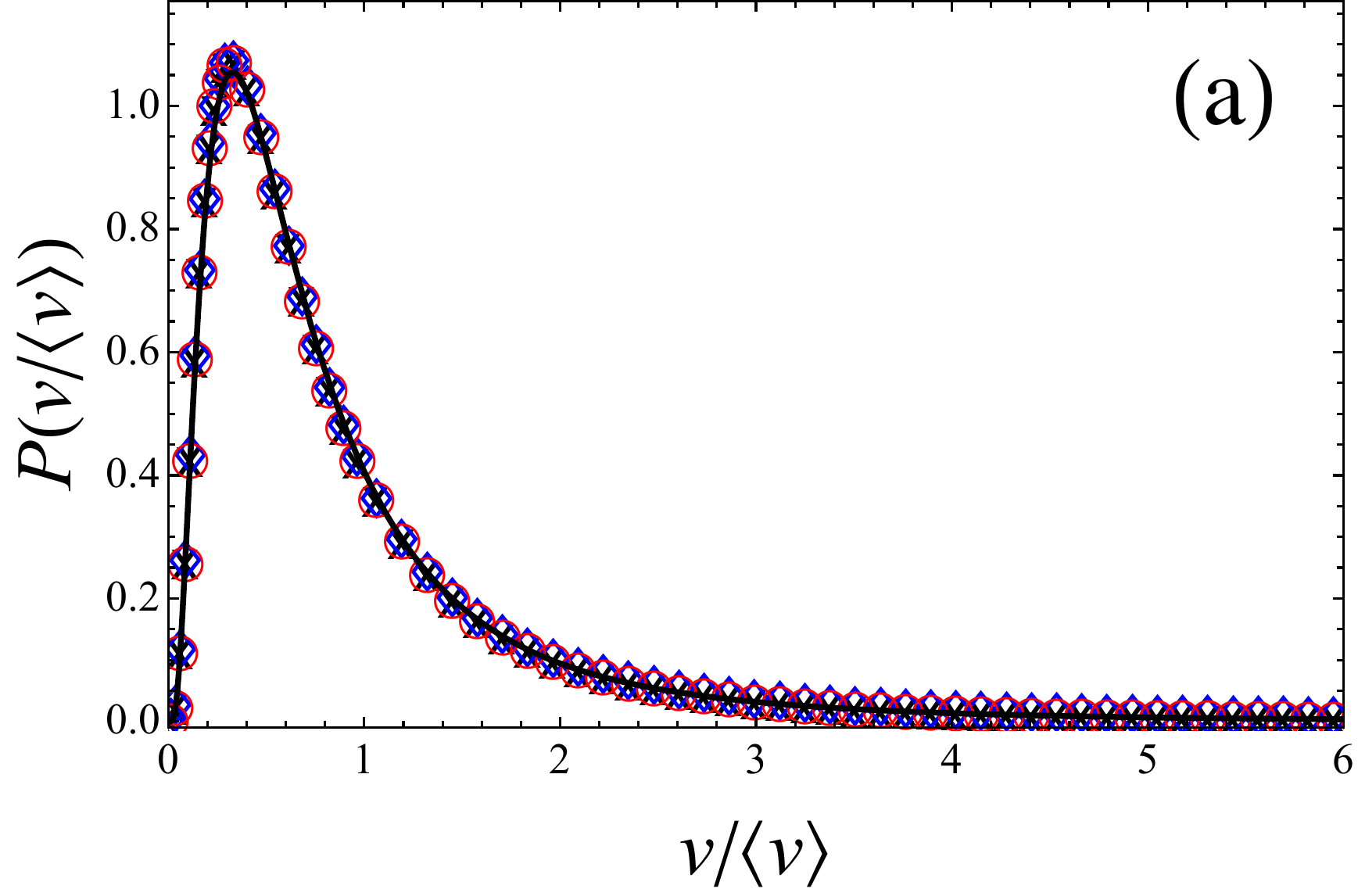}
                \includegraphics[width=0.45\textwidth]{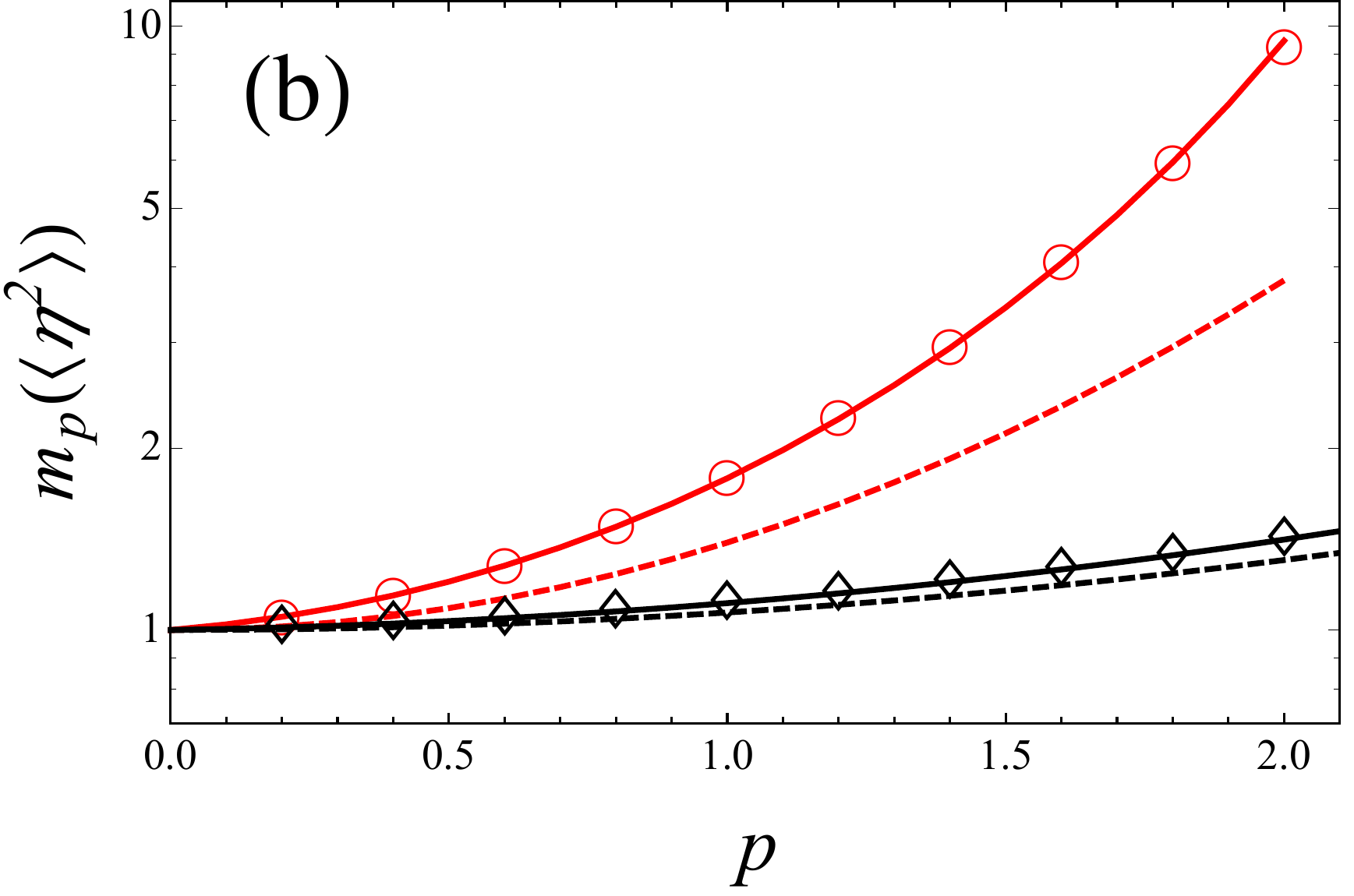}
\end{center}
\caption{ Stable behavior of the cell size in the adder model. Panel {\bf (a)}: distribution of the cell size collapse when normalized by the mean. The symbols present different $\Delta$s: $\Delta=0.1$ (circles \textcolor{red}{$\bigcirc$} ),  $\Delta=1.9$ (dimonds \textcolor{blue}{$\Diamond$} and $\Delta=5.0$ (crosses \textcolor{black}{$\times$}). The thick line is the analytic solution. Panel {\bf (b)} presents the behavior of the moments of the cell size for $\langle \eta^2 \rangle = 0.5$  (circles \textcolor{red}{$\bigcirc$}) and $\langle \eta^2 \rangle = 0.1$ (squares \textcolor{black}{$\Box$}), $\Delta$ equals $1.0$ for both cases. Thick lines represent the analytic solution for the adder model and dashed lines are solutions of the linearized adder model.}
\label{figcomparing02}
\end{figure}
%%%%%%%%%%%%%%%%%%%%%%%%%%%%%%%%%%%%%%%%%%%%%%%%%

The comparison between the distribution $P(a)$, as given in Eq.~(\ref{boltzmn}), for the three model $g(\cdot)$'s and direct simulation of the SM, Eq. (\ref{basic02}), is shown in Fig. \ref{figcomparing}. Very good agreement is observed not only at the center of the distribution but also for the tail behavior. 

The effective energy $H(a)$ in Eq.~(\ref{adderEnrg}) can be written as a function of two variables: $\exp(a)/\Delta=v/\Delta$ and $\langle \eta^2 \rangle$. This fact is sufficient to establish a separation of variables for the moments of $v$, i.e., the cell size. Each moment has the form $\langle v^p \rangle = m_p(\langle \eta^2 \rangle)\Delta^p$, where is $m_p(\dots)$ is some function that depends on the moment power $p$. Moreover, any rescaling of $v$ by $A(\langle \eta^2 \rangle)\Delta$ will produce a distribution that is independent of $\Delta$ and  a ``distribution collapse" will occur (for fixed $\langle \eta^2\rangle$), as shown in Fig.~\ref{figcomparing02}. 
The power-law dependence of the moments on $\Delta$ and separation of variables can be viewed as specific  manifestations of Taylor's law, i.e.,  $\langle v^2 \rangle \propto  \langle v \rangle^2$~\cite{Kendal,Cohen01,Giommeto2}. This phenomena was recorded for many physical/biological/ecological situations, and specifically was observed for protein number expression in {\em{E. coli}}~\cite{Xie,HSalman,Naama1}.

While we utilized a continuum approximation for the SM, a common practice is a linearization of the map around the fixed point. Specifically, for the ``adder" scenario, the linearization of $g(a)$ around $\ln(\Delta)$ produces a linear map $a_{n+1}=a_n-1/2[a_n-\ln(\Delta)]+\eta_n$. This map produces a quadratic effective energy with a minimum at $a=\ln(\Delta)$. The linearized adder effective energy is shifted with respect to the adder scenario. This fact can be observed from the location of maximum of $vP(v)$, which is located at $v=\Delta$ for the linearized adder and at $v\sim(1+\gamma)\Delta$ (where $\gamma=6\langle \eta^2 \rangle/(27+\langle \eta^2 \rangle)$) for the adder case. The presence of non-linearity and multiplicative noise is what is responsible for this shift. Not only the location of the minimum but also the shape of the effective energy is quite different, with a linear divergence of the adder as opposed to a quadratic divergence of the linearized adder. This shows itself quite strongly in the tails and higher moments of the distribution, which strongly deviate from the linearized versions as the moment power grows (see Fig.~\ref{figcomparing02}).

We have so far addressed the properties of  three specific size control scenarios using our continuum approximation of the stochastic map, characterizing the properties of the stable distributions and emergent features like the power-law behavior and discrepancies between the full and linearized size-control model.
An additional consequence of our formalism is that  the form of the effective energy in Eq.~(\ref{energy}) imposes bounds that must be satisfied by any possible size control scenario. 
First, the effective force $g(\cdot)$ must be negative for large positive values of $\ln(v)$ and positive for sufficiently large negative values of $\ln(v)$. In general it means that the effective force must be restoring, i.e. particle connected to a spring that perfectly describes the linearized adder.
Second, even when the first condition is fulfilled, due to the presence of $-g^2(a)/2$ in the effective energy the behavior of $g(a)$ in the limits $a\to\pm\infty$ is restricted. Since the  size distribution is assumed to be stable, the term $-2\int g(a)\,da$ must be dominant, restricting the growth of $g(a)$ to be linear at most. 
Basically it means that the size control, $g(a)$, is bounded between some constant value and linear growth as $a\to \pm \infty$.  
Specifically, the adder scenario fulfills the first and second conditions. What is important to notice is that for large values of $v$ the adder scenario is the least restrictive, as $g(a)\to -\ln(2)$ as $a\to\infty$.
What is meant by ``least restrictive" is that the minimal effective force is applied in order to stabilize the bacteria size.  When treating small sizes, the adder scenario is much more restrictive and a maximal effective force is applied for stabilization. From this discussion it becomes clear that the least restrictive scenario is when a minimal effective force is applied for both large and small cell sizes $v$. For large sizes, as we noted, it is an adder-like scenario. For small sizes, the minimal force ($g(a)$) is a positive constant, i.e., a timer scenario. 
It is then a mixed scenario of timer for small sizes and adder for large sizes that is a minimal scenario capable of stabilizing cell size. While we can't claim that the cell must prefer such a mechanism in order to minimize the effective energy invested in controlling the size, it is still very encouraging that exactly such a ``mixer" mechanism was very recently spotted for {\it C. crescentus}~\cite{Mixer}.

An additional constraint that must be satisfied by any growth scenario is the distribution collapse due to rescaling. From the form of the effective energy in Eq.~(\ref{energy}) and the growth scenario $g(a)$ we can conclude that for any 
growth scenario that can be written as some function of $a-f(\lambda_1,\lambda_2,...,\lambda_n)$ ($\lambda_i$ are some parameters of the scenario) a distribution collapse can be observed. If the different distributions were created by changing only the parameters $\lambda_i$ a collapse of the distributions will occur if the size $v$ is normalized by $\langle v \rangle$. This effect is dictated by the form of $H(a)$ in Eq.~(\ref{energy}) and the fact that $\langle v \rangle \propto \exp[f(\lambda_1,\lambda_2,...,\lambda_n)]$. This effect can be used as an examination of which parameters were changed under various experimental conditions, when non-collapsing distributions will have changes to parameters that are not in the set $\lambda_1,\ldots,\lambda_n$. For example, in the case of the adder scenario the collapse will occur only if $\Delta$ is changed, while changing the noise strength $\langle \eta^2 \rangle$ will produce a distribution that does not   ``collapse" under rescaling. 

%In this manuscript we have described how to address different  phenomenological scenarios of cell size control during exponential growth. We utilized our previously developed approximation in order to properly describe, and restrict, various cell control scenarios under stable conditions.   
An intriguing point is why in living systems such simple scenarios for cell-size control appear. While it is hard to address this question from a molecular perspective, the treatment of cell-size control as a map permits quite general statements. Two important points must always be satisfied, {\bf(a)} Existence of a fixed point for the map and {\bf(b)} Stability with respect to noise fluctuations. Any molecular cell-size control mechanism must eventually satisfy those restrictions. Our previously developed approximation of stochastic maps allows us to describe these restrictions in the language of physics using such terms as potential and force. General bounds are then naturally obtained  in terms of properties of the force/potential.

An additional advantage of the  ideas of effective force and energy for cellular growth and devision presented herein is the the possibility of treating static and time-dependent perturbations. This will allow for the theory to deal with with experiments that take the ``top down" approach and try to affect (and study) the single cell behavior as a whole by application of external macroscopic perturbations, e.g., chemical perturbation spectroscopy~\cite{Yihan}.

\bibliographystyle{apsrev4-1} % Tell bibtex which bibliography style to use
\bibliography{./maplit} % Tell bibtex which .bib file to use (this one is some example file in TexLive's file tree)

\end{document}